\documentclass[12pt]{spieman}

\usepackage[]{graphicx}
\usepackage{setspace}
\usepackage{tocloft}

\title{
A simple  digital system for tuning and long-term frequency 
stabilisation of a CW Ti:Sapphire laser}

\author{I.~I.~Beterov,\supscr{a,b} A.~Markovski,\supscr{c,d} S.~M.~Kobtsev,\supscr{b,e} E.~A.~Yakshina,\supscr{a,b,f} V.~M.~Entin,\supscr{a}D.~B.~Tretyakov,\supscr{a} V.~I.~Baraulya,\supscr{b,e} I.~I.~Ryabtsev\supscr{a,b,f}}

\affiliation{\supscrsm{a}Rzhanov Institute of Semiconductor Physics SB RAS, 630090 Novosibirsk, Russia\\
\supscrsm{b}Novosibirsk State University, 630090 Novosibirsk, Russia\\
\supscrsm{c}University of Latvia, LV-1002 Riga, Latvia\\
\supscrsm{d}Technical University of Sofia, Faculty of Automatics, 1000 Sofia, Bulgaria\\
\supscrsm{e}"Tekhnoscan - Lab" LLC, 630090, Novosibirsk, Russia\\
\supscrsm{f}Russian Quantum Center, Skolkovo, Moscow Reg., 143025, Russia}

\cftpagenumbersoff{figure}
\cftpagenumbersoff{table} 
\begin{document} 
\maketitle

\begin{abstract}
We have implemented a simple digital system for long-term 
frequency stabilisation and locking to an arbitrary wavelength of the 
single-frequency ring CW Ti:Sapphire laser. This system is built using two 
confocal Fabry-P\'erot cavities, one of which is used to narrow short-term 
linewidth of the laser and the other to improve long-term stability of 
the laser frequency. The length of the second cavity is stabilised using the 
radiation from an external-cavity diode laser locked to an atomic transition. 
Our system is an improvement of a commercial Tekhnoscan laser lock. This 
system has been successfully used in our experiments on high-resolution laser 
spectroscopy of ultracold rubidium Rydberg atoms.
\end{abstract}


{\noindent \footnotesize{\bf Address all correspondence to}: Ilya I. Beterov, Rzhanov Institute of Semiconductor Physics SB RAS, prosp. Lavrentyeva, 13 630090 Novosibirsk, Russia; Tel: +7 383-333-24-08; Fax: +7 383-333-27-71; 
E-mail:  \linkable{beterov@isp.nsc.ru}}

\begin{spacing}{2}

\section{Introduction}

Laser cooling and multiphoton spectroscopy of atoms and molecules commonly require 
the use of narrow-band frequency stabilised lasers working at various wavelengths. 
Single-frequency CW Ti:Sapphire lasers with ring cavities are advantageous 
for laser spectroscopy due to their high power in single frequency and 
tuneability over a broad spectral range. A variety of techniques have been 
developed for laser frequency stabilisation needed in many applications in 
optics and spectroscopy. Typically, the laser output is locked to a highly 
stable reference cavity or to an atomic transition by means of fringe-side 
locking~\cite{Barger1973} or most widely used Pound-Drever-Hall (PDH) technique~\cite{Black2001,Drever1983}. 
Highly-stable Fabry-P\'erot cavities commonly use ultra-low expansion (ULE) glass spacers and are 
temperature-stabilised and kept in vacuum in order to avoid drifts of the 
resonant frequency induced by temperature and air pressure variations~\cite{PasqualeMaddaloni2013, Nature2011, Nature2012}. 
In such systems, the mirrors cannot be placed on piezo-ceramic 
transducers (PZT), which feature large frequency drifts of tens of MHz per 
hour~\cite{Low2012}. Therefore, the laser frequency can only be stabilized at a series of fixed values, separated by the free spectral range of the cavity. To overcome this obstacle it is possible to shift the laser frequency  using an external acousto-optical modulator or sideband locking technique with electro-optical modulator~\cite{Thorpe2008}.
Locking the laser to an atomic transition via saturation 
spectroscopy is commonly used in experiments on atomic spectroscopy and 
laser cooling. However, this technique is difficult to apply in the case 
when the desirable laser wavelength does not match the wavelength of 
available atomic transitions. The same difficulty occurs when the reference cavity is locked to a highly stable reference laser.
Another recently developed method is based on 
the use of a high-precision wavelength meter, which also requires additional 
reference laser to achieve high accuracy~\cite{Kobtsev2007}.

One of the common experimental demands is multistep excitation of ultracold atoms which requires locking of the laser 
to a transition between excited atomic states. For two-step excitation it is possible to use the resonances of electromagnetically-induced transparency in a gas cell to lock the lasers~\cite{Abel2009}. However, generalization of this scheme for three-step laser excitation is technically challenging due to small dipole moments of transitions to Rydberg states and necessity to combine three laser beams in a vapor cell.

We have developed a relatively simple and inexpensive method relying on the
digital measurement of the frequency difference between the output of a Ti:Sapphire laser and that of a highly stable diode laser  using an auxiliary Fabry-P\'erot 
interferometer. The diode lasers which are locked to an atomic transition between the ground and first excited states of alkali-metal atoms are commonly used in laser cooling experiments. It is important that a single reference laser can be used to lock other lasers to the arbitrary wavelengths.

Essentially, in our locking system the 
time delay is measured between emergence of transmission peaks coming from 
the radiation of the Ti:Sapphire laser and from that of the highly stable 
diode laser as the Fabry-P\'erot cavity is scanned.

This method has been successfully implemented in several previous works 
\cite{Jaffe1993, Zhao1998, Matsubara2005, Seymour-Smith2010}. A rather complicated 
analogue electronic detection of transmission peaks in the scanning cavity 
was first implemented in Ref.~\cite{Jaffe1993}. A digital integrator was used to 
provide the feedback signal to the laser. The frequency drift of the laser did 
not exceed 1~MHz compared to the reference frequency-stabilised He-Ne laser. 
In the later works, the electronics for laser stabilisation was substantially 
simplified by introduction of ADC/DAC modules with a computer control 
\cite{Zhao1998, Matsubara2005, Seymour-Smith2010}. In Ref.~\cite{Zhao1998} the interference 
filters were used to individually measure the transmission of multiple laser 
beams through the cavity, so that several lasers could be locked 
simultaneously. In Ref.~\cite{Matsubara2005} the scanning rate was increased to 1~kHz 
compared to 200~Hz in Ref.~\cite{Zhao1998} and the transmission signal was sampled 
at 4~MHz. Increased scanning rate permits frequency locking in a wider 
bandwidth, but requires faster and more expensive ADC conversion and data analysis. In Ref.~\cite{Seymour-Smith2010} the scanning frequency was further raised to 
3~kHz and the feedback signal was generated using an analogue peak detector 
and a fast programmable microcontroller.
\begin{figure}[!t]
\includegraphics[width=\columnwidth]{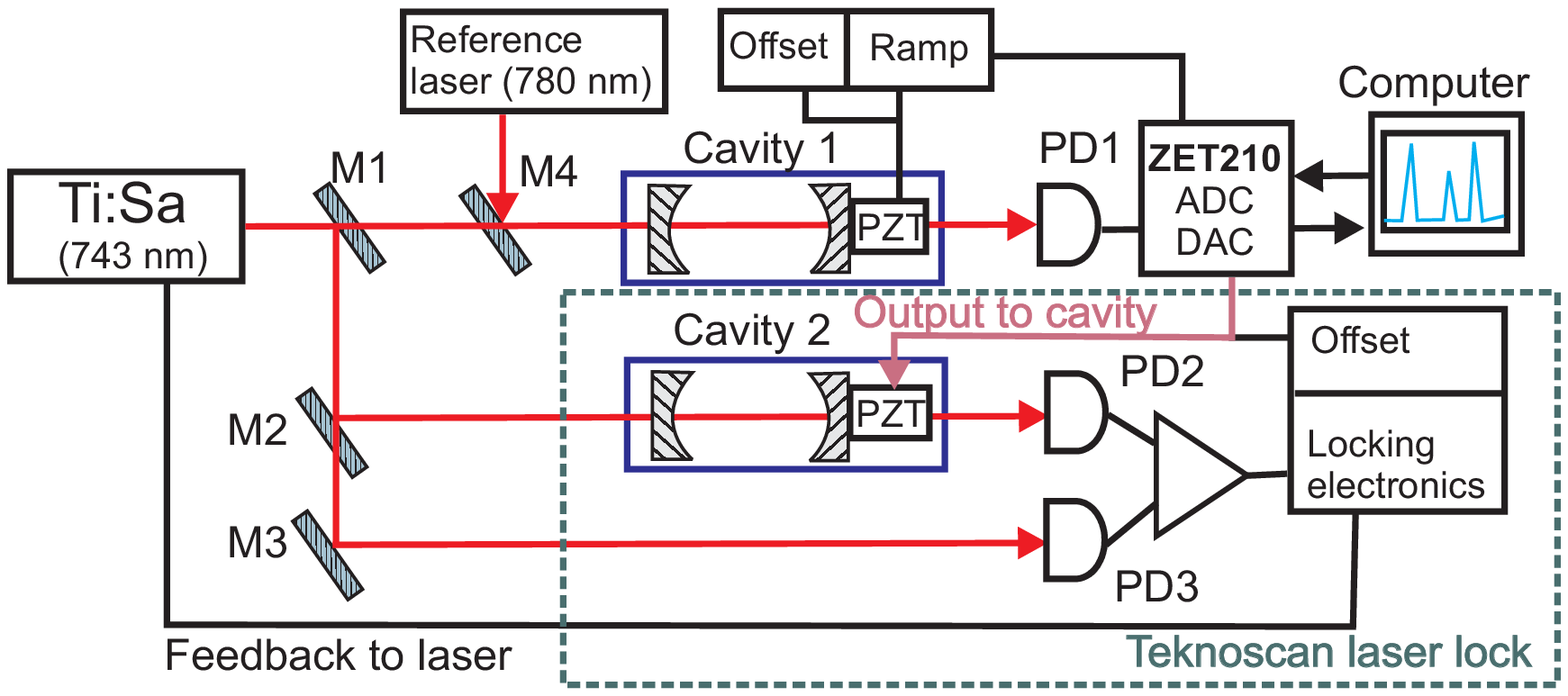}
\vspace{-.5cm}
\caption{
\label{Scheme}(Color online).
Scheme of the experimental setup. The Ti:Sapphire 
laser is locked to cavity~2 using side-fringe locking technique. Cavity 1 is 
scanned at 200~Hz. The transmission peaks from two lasers are detected on 
PD1, and recorded using ADC, and analysed on a computer. The feedback signal 
from DAC is sent to the PZT of cavity~2. The part of the experimental setup which represents the commercial Teknoscan laser lock is shown in the rectangular box.
}
\end{figure}

Our approach relies on two confocal cavities for line narrowing and 
long-term frequency stabilisation of the laser. The  confocal 
reference cavity with the mirrors separated by a 10-cm invar spacer is more stable 
than a long (approximately 70~cm) laser resonator. We use 
fast analogue locking electronics manufactured by the Tekhnoscan Company 
(Novosibirsk, Russia) to lock the laser on the reference cavity, and then 
we only compensate for the long-term drift of the optical length of the 
reference cavity. This technique allows us to avoid the necessity of using 
high-speed electronics for peak detection and data analysis, comparing to the previous works~\cite{Matsubara2005, Seymour-Smith2010}. We use a relatively cheap ZetLab$^{TM}$ Zet210 ADC/DAC module (Russia) sampled at 400~kHz for data acquisition and generation of the feedback output voltage, and National Instruments LabView$^{TM}$ for data analysis.

\section{Experiment}

The scheme of our experimental setup is shown in Fig.~\ref{Scheme}. The radiation of 
the Ti:Sapphire laser is split by semi-transparent mirror M1 and directed to 
confocal cavities 1 and 2. The intensity of the light transmitted through the 
cavities is measured with photodiodes PD1 and PD2. Semi-transparent mirror 
M2 is used to create a reference laser beam for the side-fringe locking of the 
Ti:Sapphire laser to a mode of cavity 2 by measuring the difference between 
the intensities on photodiodes PD2 and PD3 and sending the feedback signal 
to the laser. This is a part of the commercial Tekhnoscan laser lock. 
Cavity~2 is a confocal Fabry-P\'erot resonator with a 10-cm 
invar spacer and one of the mirrors mounted on a PZT. The commercially available 
locking electronics by Tekhnoscan reduces the short-term linewidth of laser 
radiation to less than 10~kHz. By applying an offset voltage to the PZT of 
cavity~2, it is possible to tune the laser frequency within the free 
spectral range of the cavity, which is 750~MHz. The laser frequency drift caused 
by temperature and air pressure variations, as well as the drift of the PZT, is 
more than 30~MHz per hour even for the cavity with temperature stabilisation.  
This does not meet our requirements for experiments with cold Rydberg 
atoms where resonance widths at laser excitation are less than 5~MHz, and 
we need to keep the laser on resonance with the atomic transition for tens of minutes.

In order to compensate for this drift, we combine on semi-transparent mirror 
M4 the radiation of the Ti:Sapphire laser, which is already locked to cavity~2, 
with the radiation of the reference external-cavity diode laser with wavelength 780~nm, 
locked to an atomic transition in a rubidium vapour cell via Pound-Drever-Hall technique~\cite{Black2001,Drever1983}.
The linewidth of this laser was measured by observing the beat spectrum of two identical lasers averaged during 100~ms, and was found to be less than 1~MHz.
The radiation of both lasers is send to cavity 1 which is 
scanned by a triangular signal at 200~Hz from a GW Instek DDS function 
generator SFG-2004$^{TM}$. Low-voltage output of the function generator is 
amplified with a home-built high-voltage amplifier, which drives the PZT. 
This amplifier also provides additional offset DC voltage to the PZT.

The scanning cavity~1 is a confocal Fabry-P\'erot resonator with a 12.5-cm invar 
spacer corresponding to a free spectral range of $c/4L = 600$~MHz. The 
cavity can be temperature stabilised, however this does not significantly 
change the performance of the system. The signal from photodiode PD1, which 
has a bandwidth of 10~MHz, is measured using a commercial ZetLab
ZET~210$^{TM}$ ADC/DAC module with maximum sampling frequency of 400~kHz.

\begin{figure}[!t]
\includegraphics[width=\columnwidth]{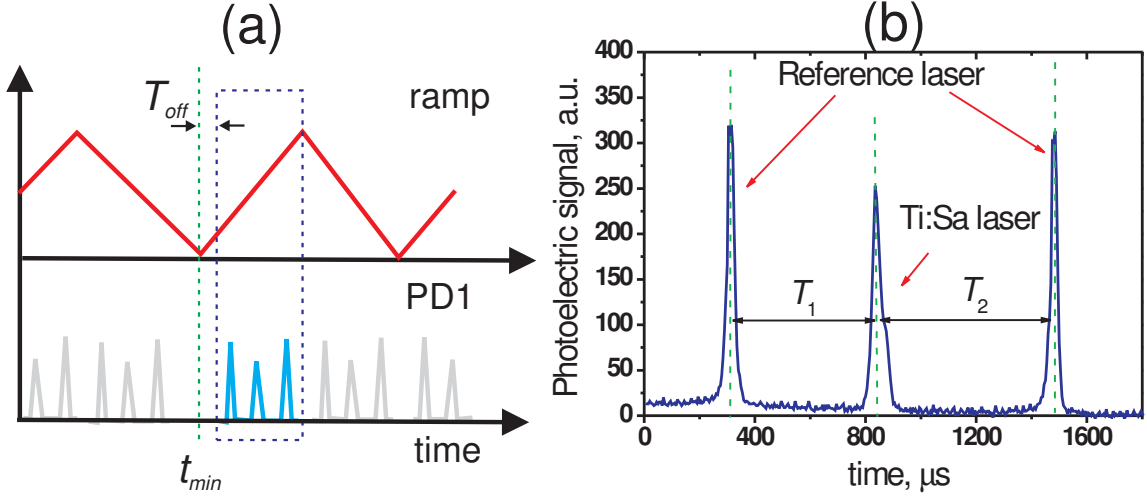}
\vspace{-.5cm}
\caption{
\label{Timing}(Color online).
(a) Timing diagram of the measurement. The ramp signal is used 
for synchronisation. The first minimum of the ramp signal and the 
measurement time bin are separated by offset time 
$T_{\rm off}$; (b) Time-dependent photoelectric 
signal on PD1.
}
\end{figure}

As the signal from photodiode PD1 can only be measured continuously during 
the period defined by the ADC buffer size of 4096 bytes, we have to use the 
signal from the function generator, which is measured by a second ADC 
channel, for synchronization of the data acquisition process. We read the 
data from the ADC buffer, which contains at least two periods of the 
synchronisation signal. Then we find the time of the first minimum of the 
synchronization signal $t_{\rm min}$ and extract data from a time bin 
defined by the specified width and offset $T_{\rm off}$ as shown in 
Fig.~\ref{Timing}(a). Adjustment of the DC offset voltage at the PZT of cavity~1 can be 
used to tune the positions of the peaks within the window without 
interrupting the frequency locking process. This is necessary to move the 
transmission peaks away from the turning points of the ramp signal and to 
compensate for temperature drift of the cavity in order to keep all the 
peaks within the measurement window.

The measured time-dependent photoelectric signal on PD1, which is 
proportional to the transmission of cavity~1, is plotted in Fig.~\ref{Timing}(b). 
We measure the time intervals 
$T_1$ and $T_2$ between two peaks from the reference 
laser and the peak from the Ti:Sapphire laser using the Peak Detect virtual 
instrument implemented in National Instruments Labview$^{TM}$, which uses quadratic fit approximation for accurate determination of the peak centers. Then we calculate the ratio:

\begin{equation}
\label{eq1}
R = \frac{{T_1} }{{T_1 + T_2} }
\end{equation}

\noindent  This ratio depends linearly on the frequency of the Ti:Sapphire laser and 
ranges between 0 and 1 depending on the relation between the frequencies of 
the reference laser and Ti:Sapphire laser. We apply a 3-point average filter to an array of measured ratios.

Initially, the Ti:Sapphire laser is locked to cavity~2. By changing the offset 
voltage on the PZT of this cavity we can tune the laser to any desired 
wavelength which define the initial ratio $R_0$ of time 
intervals between the transmission peaks of cavity~1. For long-term 
stability, we need to minimize the variation $\Delta R = R - R_0 $. This 
is achieved by applying an additional voltage to the PZT of cavity 2. The 
feedback signal is calculated using National Instruments Labview$^{TM}$ PID 
toolkit and then converted to voltage via the DAC of the Zetlab ZET~210$^{TM}$ module. 

To tune the laser locked to cavity~1, we can manually change the initially 
set value $R_0$ in the control program during the experiment. The locking 
system then drives the laser to a newly defined locking point.

\section{System performance}

\begin{figure}[!t]
\includegraphics[width=\columnwidth]{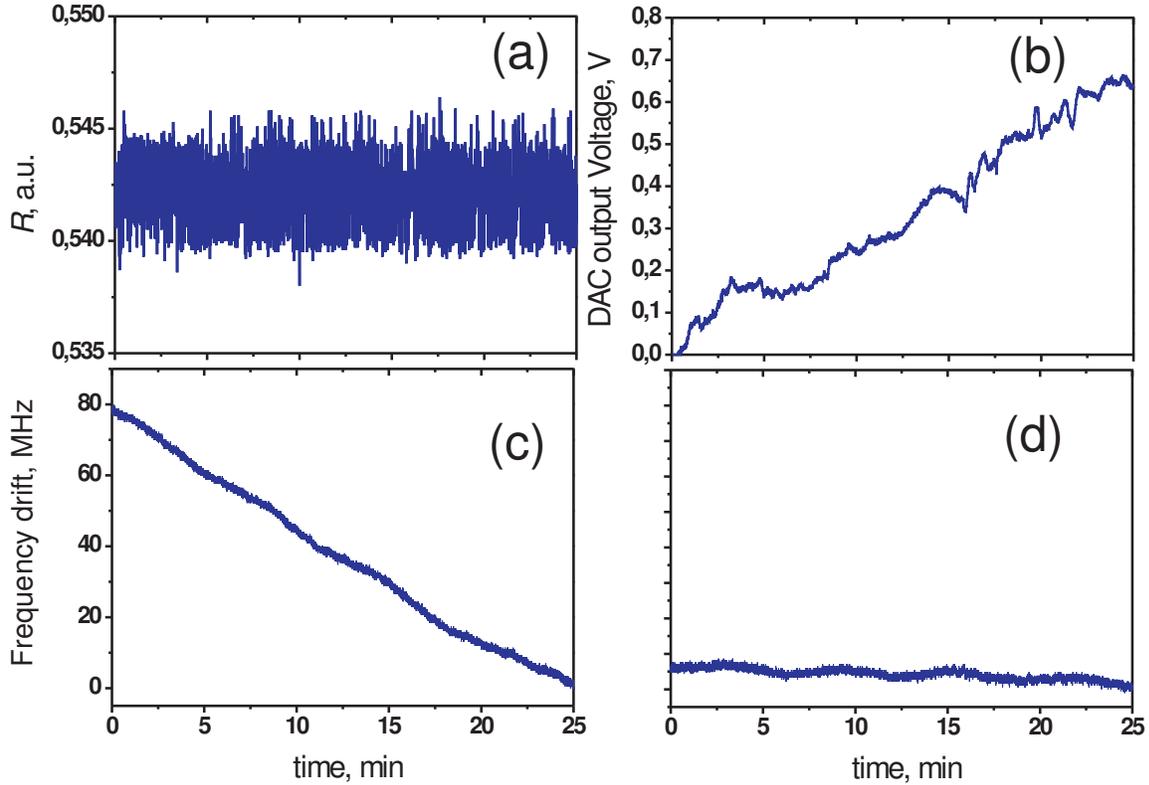}
\vspace{-.5cm}
\caption{
\label{Performance}(Color online).
(a) Measured time trace of ratio $R$ of time intervals between the transmission 
peaks; (b) time trace of DAC output voltage; (c) measured laser frequency drift with 
the Ti:Sapphire laser locked to cavity~2 only (without temperature 
stablization of the cavity); (d) measured laser frequency drift with the Ti:Sapphire laser locked to 
both cavity~1 and cavity~2. The absolute long-term accuracy of the wavemeter is 200~MHz, but the relative accuracy is much better.
}
\end{figure}

The results of measurement of the ratio of the time intervals \textit{R} 
during approximately 20 minutes are presented in Fig.~\ref{Performance}(a). The standard 
deviation of $R$ was 0.00115 which corresponds to the error of 0.7~MHz 
in the determination of the laser frequency. This result is consistent with 
the previous measurements~\cite{Jaffe1993, Zhao1998, Matsubara2005, 
Seymour-Smith2010}. The output voltage of DAC, shown in Fig.~\ref{Performance}(b), was 
automatically increased by the locking system from zero to around 0.7~V 
during the measurement due to the temperature and air pressure variations which 
affect the optical length of cavity~2, and due to the drift of the PZT of 
cavity~2. 

We have also measured the wavelength of Ti:Sapphire laser output using High 
Finesse$^{TM}$ WS6 wavemeter. Although this model has limited long-term accuracy 
(200~MHz error in absolute value), we have used it to study the short-term fluctuations 
and drifts of the laser frequency. In the mode when the laser was locked to 
cavity~2 only without temperature stabilisation of the cavity the drift was 
around 3~MHz per minute [see Fig.~\ref{Performance}(c)]. When the laser was also locked to 
cavity~1 the measured drifts were reduced to~200~kHz per minute [see 
Fig.~\ref{Performance}(d)] which can be attributed to the temperature drift of the wavemeter. 
Another possible source of error could be the drift of the reference laser, but we have not observed it in the beat spectrum of two identical reference lasers locked to 780~nm. 

Our approach is close to the laser locking system described in Ref.~\cite{Zhao1998} where low-frequency data acquisition at 400~kHz had been used along with fully digital data analysis. However, it has been noted in Ref.~\cite{Zhao1998} that their system was unable to compensate for acoustic noises, and it has been proposed to use an additional stable cavity and analog electronics for linewidth narrowing. In our work the laser is locked to cavity~2 by analog electronics via fringe-side locking, which removes the acoustic noises and provides less than 10~kHz rms laser linewidth relative to cavity~2.  Our digital locking system compensates only for slow drifts of cavity~2. These are the main differences from the previous work~\cite{Zhao1998}.

The limiting factors for the system performance are rather low scanning rate 
(200~Hz), finesse of cavity~2 ($F\sim 50$), and stability of the 
reference laser. Higher scanning rate requires faster sampling, which was 
not possible with the ADC we used. However, the peak positions fluctuate from scan to 
scan even for a stabilised laser due to the nonlinearities and hysteresis of 
the PZT~\cite{Zhao1998}. Therefore we believe that increase of the scanning rate will not substantially improve the performance of our system, and we need to compensate only for slow temperature drifts of cavity~2 and air pressure variations. There is no need to further increase the bandwidth of the locking system by use of high-speed digital electronics, in contrast to previous works~\cite{Matsubara2005, Seymour-Smith2010}. Besides that, for a faster lock, the fluctuations of the peak positions can substantially increase the linewidth of the laser.

\begin{figure}[!t]
\includegraphics[width=\columnwidth]{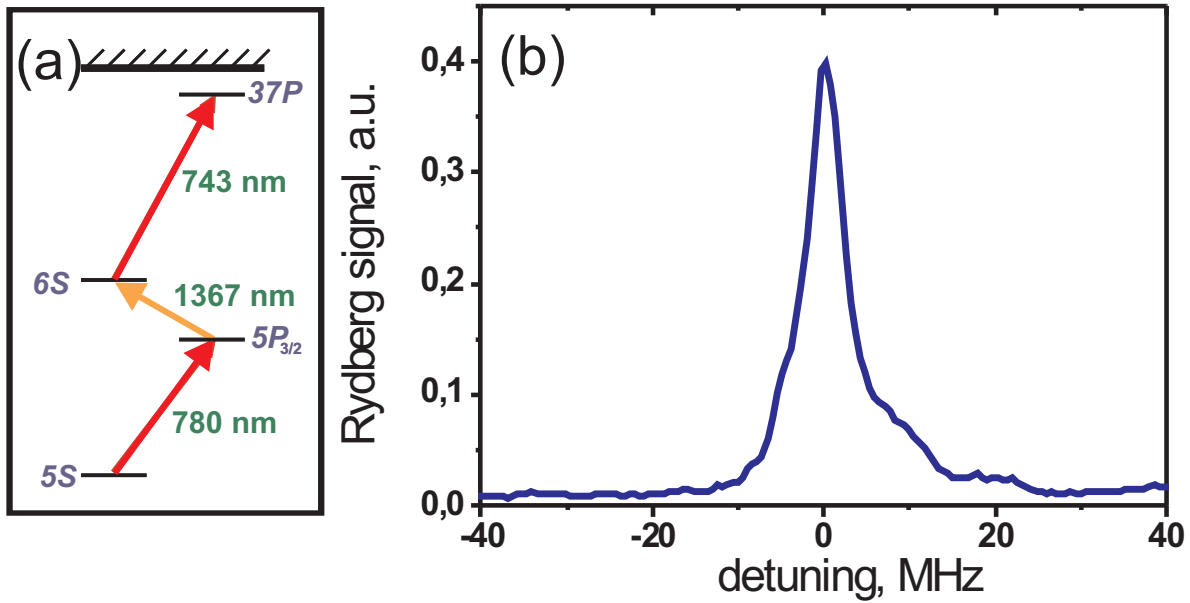}
\vspace{-.5cm}
\caption{
\label{Rydberg}(Color online).
(a) Scheme of the three-photon excitation of rubidium Rydberg atoms; (b) A typical spectrum of laser excitation 
of cold rubidium 37$P$ Rydberg atoms obtained by scanning of the Ti:Sapphire laser working at 743~nm. 
}
\end{figure}

The system performance was good enough to compensate for the long-term drift of 
cavity~2. We have used this system to lock the Ti:Sapphire laser at 743~nm 
for three-step laser excitation of cold rubidium atoms into the Rydberg 
states, as shown in Fig.~\ref{Rydberg}(a)~\cite{Tretyakov2014, Entin2013}. We used a cooling 780~nm laser diode as a reference source. The typical spectrum of laser excitation of 37$P$ Rydberg rubidium atoms obtained by scanning of the Ti:Sapphire laser by linear ramp voltage applied to cavity~2 is shown in Fig.~\ref{Rydberg}(b). The width of the atomic resonance in our experiment was around 5~MHz. Therefore the stability of our laser lock was good enough for Rydberg excitation of ultracold rubidium atoms.

\acknowledgments 

This work was supported by RFBR grant No. 14-02-00680, 
by the Russian Academy of Sciences, by EU FP7 IRSES Project "COLIMA", and 
by the Russian Quantum Center.



\vspace{2ex}\noindent{\bf Ilya I. Beterov} is a senior researcher at the Rzhanov Institute of Semiconductor Physics, Siberian Branch of Russian Academy of Sciences and lecturer at Novosibirsk State University. He received Ph.D.  degree from the Institute of Automation and Electrometry Siberian Branch of Russian Academy of Sciences in 2008. His research area is quantum information with Rydberg atoms. He is the author of more than 30 publications in scientific journals. 
\vspace{1ex}

\vspace{2ex}\noindent{\bf Asparuh G. Markovski}  works at the Technical University of Sofia, Bulgaria, at the Faculty of Automation. He received his Ph. D. from the same University in 2004. His major research interests are in the area of Robust Control, Low-cost means for automation, Optimization in the Control Theory.
\vspace{1ex}

\vspace{2ex}\noindent{\bf Sergey Kobtsev}  is a head of Division of Laser Physics and Innovative Technologies at Novosibirsk State University. He received PhD (1992) and Doctor of Sciences (2010) degrees from the Institute of Automation and Electrometry, Siberian Branch of the Russian Academy of Sciences. His research areas include quantum electronics, fiber and nonlinear optics, high-precision and tunable laser systems. He is the author of more than 200 publications in scientific journals, conference proceedings, and books.
\vspace{1ex}

\vspace{2ex}\noindent{\bf Elena A. Yakshina } is a Ph.D. student of Novosibirsk State University. Her major research interest is laser spectroscopy of Rydberg atoms.
\vspace{1ex}

\vspace{2ex}\noindent{\bf  Vasily M. Entin} is a senior researcher at the Rzhanov Institute of Semiconductor Physics, Siberian Branch of Russian Academy of Sciences. He received his Ph.D. from the Institute of Automation and Electrometry, Siberian Branch of the Russian Academy of Sciences in 2006. His major research interests are laser cooling and laser spectroscopy. He is the author of more than 30 publications in scientific journals.
\vspace{1ex}

\vspace{2ex}\noindent{\bf Denis B. Tretyakov } is a researcher at the Rzhanov Institute of Semiconductor Physics, Siberian Branch of Russian Academy of Sciences. He received his Ph.D. from the Institute of Automation and Electrometry, Siberian Branch of the Russian Academy of Sciences in 2008. His major research interests are quantum information with Rydberg atoms and long-range atomic interactions. He is the author of more than 30 publications in scientific journals. 
\vspace{1ex}

\vspace{2ex}\noindent{\bf  Vladimir Baraulya} graduated in 1982 from Novosibirsk State University in quantum optics. His research activity was focused upon such areas of high importance as systems for laser cooling of atoms and molecules, frequency doublers, laser physics. 
\vspace{1ex}

\vspace{2ex}\noindent{\bf Igor I. Ryabtsev}  is a head of laboratory at Rzhanov Institute of Semiconductor Physics, Siberian Branch of Russian Academy of Sciences and lecturer at Novosibirsk State University. He received Ph.D. (1992) and Doctor of Sciences (2005) degrees from Institute of Automation and Electrometry, Siberian Branch of the Russian Academy of Sciences in 2005. His major research interests are quantum information with Rydberg atoms and long-range atomic interactions. He is the author of more than 90 publications in scientific journals.
\vspace{1ex}

\listoffigures
\listoftables

\end{spacing}

\begin{thebibliography}{10}
\newcommand{\enquote}[1]{``#1''}


\bibitem{Barger1973}
R.~L. Barger, M.~S. Sorem, and J.~L. Hall, \enquote{Frequency stabilization of
  a cw dye laser,} Appl. Phys. Lett. \textbf{22}, 573 (1973).

\bibitem{Black2001}
E.~D. Black, \enquote{An introduction to pound–drever–hall laser frequency
  stabilization,} Am. J. Phys \textbf{69}, 79 (2001).

\bibitem{Drever1983}
R.~W.~P. Drever, J.~L. Hall, F.~V. Kowalski, J.~Hough, G.~M. Ford, A.~J.
  Munley, and H.~Ward, \enquote{Laser phase and frequency stabilization using
  an optical resonator,} Appl. Phys. B \textbf{31}, 97 (1983).

\bibitem{PasqualeMaddaloni2013}
P.~Maddaloni, M.~Bellini, and P.~De~Natale, \emph{Laser-Based Measurements for
  Time and Frequency Domain Applications: A Handbook (Series in Optics and
  Optoelectronics)} (CRC Press, 2013).
	
	\bibitem{Nature2011} Y.~Y.~Jiang,	A.~D.~Ludlow,	N.~D.~Lemke,	R.~W.~Fox,	J.~A.~Sherman,	L.-S.~Ma, C.~W.~Oates, \enquote{Making optical atomic clocks more stable with $10^{−16}$-level laser stabilization,}    Nature Photonics    \textbf{5},     158   (2011).

\bibitem{Nature2012}
 T.~Kessler,	C.~Hagemann,	C.~Grebing,	T.~Legero,	U.~Sterr,	F.~Riehle,	M.~J.~Martin,	L.~Chen, J.~Ye, \enquote{A sub-40-mHz-linewidth laser based on a silicon single-crystal optical cavity,}     Nature Photonics  \textbf{6},  687 (2012).

\bibitem{Low2012}
R.~L\"ow, H.~Weimer, J.~Nipper, B.~Balewski, J. B.~Butscher, H.~B. B\"uchler,
  and T.~Pfau, \enquote{An experimental and theoretical guide to strongly
  interacting rydberg gases,} J. Phys. B \textbf{45}, 113001 (2012).


\bibitem{Thorpe2008} J.~I.~Thorpe, K.~Numata, and J.~Livas, \enquote{Laser frequency stabilization and
control through offset sideband locking to optical cavities,} Optics Express \textbf{16}, 15980 (2008).

\bibitem{Kobtsev2007}
S.~Kobtsev, S.~Kandrushin, and A.~Potekhin, \enquote{Long-term frequency
  stabilization of a continuous-wave tunable laser with the help of a precision
  wavelength meter,} Appl. Opt. \textbf{46}, 5840 (2007).

\bibitem{Abel2009}
R.~P.~Abel, A.~K.~Mohapatra, M.~G.~Bason, J.~D.~Pritchard, K.~J.~Weatherill, U.~Raitzsch and C.~S.~Adams \enquote{Laser frequency stabilization to highly excited state transitions using electromagnetically induced transparency in a cascade system}
Appl. Phys. Lett. 94 , 071107 (2009). 


\bibitem{Jaffe1993}
S.~M. Jaffe, M.~Rochon, and W.~M. Yen, \enquote{Increasing the frequency
  stability of singlefrequency lasers,} Rev. Sci. Instrum. \textbf{64}, 2475
  (1993).

\bibitem{Zhao1998}
W.~Z. Zhao, J.~E. Simsarian, L.~A. Orozco, and G.~D. Sprouse, \enquote{A
  computer-based digital feedback control of frequency drift of multiple
  lasers,} Rev. Sci. Instrum. \textbf{69}, 3737 (1998).

\bibitem{Matsubara2005}
K.~Matsubara, S.~Uetake, H.~Ito, Y.~Li, K.~Hayasaka, and M.~Hosokawa,
  \enquote{Precise frequency-drift measurement of extended-cavity diode laser
  stabilized with scanning transfer cavity,} Japanese Journal of Applied
  Physics \textbf{44}, 229 (2005).

\bibitem{Seymour-Smith2010}
N.~Seymour-Smith, P.~Blythe, M.~Keller, and W.~Lange, \enquote{Fast scanning
  cavity offset lock for laser frequency drift stabilization,} Rev. Sci.
  Instrum. \textbf{81}, 075109 (2010).

\bibitem{Tretyakov2014}
D.~B.~Tretyakov, V.~M.~Entin, E.~A.~Yakshina, I.~I.~Beterov, C.~Andreeva, and I.~I.~Ryabtsev, \enquote{Controlling the interactions of a few cold rb rydberg atoms by radiofrequency-assisted F\"orster resonances,} Phys.~Rev.~A 90, 041403(R) (2014).

\bibitem{Entin2013}
V.~M.~Entin, E.~A.~Yakshina, D.~B.~Tretyakov, I.~I.~Beterov, and I.~I.~Ryabtsev, \enquote{Spectroscopy of the three-photon laser excitation of cold
  rubidium rydberg atoms in a magneto-optical trap,} J. Exp. Theor. Phys.
  \textbf{116}, 721 (2013).

\end{thebibliography}
\end{document}